\documentclass[review]{elsarticle}

\graphicspath{{./figures/}}

\usepackage[T1]{fontenc}
\usepackage[utf8]{inputenc}
\usepackage{lmodern}
\usepackage{courier}
\usepackage{amsmath,amssymb,amsthm}
\usepackage{graphicx}
\usepackage{booktabs}
\usepackage{array}
\usepackage{tabularx}
\usepackage{etoolbox}
\usepackage{caption}

\usepackage{xurl}

\usepackage[skins,listings]{tcolorbox}
\usepackage{listings}

\usepackage{hyperref}

\AtBeginEnvironment{figure}{\linespread{1}\selectfont}
\AtBeginEnvironment{table}{\linespread{1}\selectfont}
\AtBeginEnvironment{figure*}{\linespread{1}\selectfont}
\AtBeginEnvironment{table*}{\linespread{1}\selectfont}
\makeatletter
\def\fps@figure{tp}
\def\fps@table{t}
\makeatother

\newtcblisting{promptbox}{
  enhanced,
  listing only,
  listing engine=listings,
  width=\linewidth,
  colback=black!3,
  colframe=black!30,
  boxrule=0.35pt,
  arc=1.8mm,
  outer arc=1.8mm,
  boxsep=0pt,
  left=1.5mm,
  right=1.5mm,
  top=1mm,
  bottom=1mm,
  before skip=1.5mm,
  after skip=1.5mm,
  listing options={
    basicstyle=\linespread{1}\ttfamily\fontsize{6.8pt}{7.4pt}\selectfont,
    columns=fullflexible,
    breaklines=true,
    breakatwhitespace=true,
    keepspaces=true,
    showstringspaces=false,
    aboveskip=0pt,
    belowskip=0pt,
    tabsize=2
  }
}

\journal{Expert Systems with Applications}

\biboptions{authoryear}

\hypersetup{
  hidelinks,
  pdftitle={MatExtractBench}
}

\begin{document}
\begin{frontmatter}

\title{MatExtractBench: An Entity-Conditioned Document-Level Parameter Extraction Benchmark for Materials Science Literature}

\author[aff1,aff2]{Jian He}
\ead{1732413500@qq.com}

\author[aff1,aff2]{Yao Lu}
\ead{yaolu.zjut@gmail.com}

\author[aff1,aff2]{Xin Cheng}
\ead{1731867189@qq.com}

\author[aff1,aff2]{Chenxiao Liu}
\ead{1425716445@qq.com}

\author[aff1,aff2]{Tianyi Jiang}
\ead{josieyi0319@163.com}

\author[aff5]{Feng Xu}
\ead{xufeng@zjut.edu.cn}

\author[aff3]{Ying Li}
\ead{liying@zjut.edu.cn}

\author[aff6]{Qi Xin}
\ead{xinqi@zjut.edu.cn}

\author[aff5]{Yi Dong}
\ead{dongyi@zjenergy.com.cn}

\author[aff4]{Peng Liu}
\ead{peng.liu.1@zju.edu.cn}

\author[aff1,aff2]{Shanqing Yu\corref{cor1}}
\ead{yushanqing@zjut.edu.cn}

\author[aff1,aff2]{Qi Xuan}
\ead{xuanqi@zjut.edu.cn}

\cortext[cor1]{Corresponding author.}

\address[aff1]{Institute of Cyberspace Security, Zhejiang University of Technology, Hangzhou 310023, China}

\address[aff2]{Binjiang Institute of Artificial Intelligence, Zhejiang University of Technology, Hangzhou 310056, China}

\address[aff3]{Institute of Industrial Catalysis, Zhejiang University of Technology, Hangzhou 310032, China}

\address[aff4]{State Key Laboratory of Clean Energy Utilization and Institute for Carbon Neutrality, Zhejiang University, Hangzhou 310027, China}

\address[aff5]{Zhejiang Baima Lake Laboratory Co., Ltd., Hangzhou 310051, China}

\address[aff6]{College of Energy and Carbon Neutrality, Zhejiang University of Technology, Hangzhou 310014, China}
\begin{abstract}
Materials science literature provides valuable information on material composition, experimental conditions, and performance properties. However, structured parameter extraction remains challenging due to dispersed evidence and complex associations between material instances and experimental conditions. To address these challenges, this study introduces MatExtractBench, an entity-conditioned document-level parameter extraction benchmark covering nanozymes and perovskite solar cells (PSCs). Given predefined target material instances and parameter schemas, the benchmark evaluates structured parameter recovery using a RagFlow-based paper-level isolated workflow, in which each nanozyme paper is processed individually, while the main article and supplementary materials of each PSC paper form a single evidence group. After extracting and saving predictions for all target instances, the corresponding knowledge-base content is removed. Experiments evaluate locally deployed models and two API-based models across four extraction configurations. The main workflow achieves cross-model mean F1 scores of 21.64\% and 26.85\% on the nanozyme and PSC datasets, respectively, outperforming cumulative-knowledge-base retrieval by 14.22 and 25.09 percentage points. Qwen-Max achieves the highest F1 scores of 36.86\% and 52.45\%, respectively. These findings highlight the importance of evidence organization and input conditions for reliable materials parameter extraction. MatExtractBench provides a unified benchmark for instance-aligned, field-level evaluation across heterogeneous materials literature.
\end{abstract}
\end{frontmatter}

\raggedbottom

\section{Introduction}
Data-driven materials research relies on computable knowledge derived from experiments, simulations, and the literature. However, a substantial amount of materials information remains scattered in unstructured or semi-structured form across article texts, tables, figures, and supplementary materials. Relationships among material composition, preparation conditions, structural characteristics, and performance metrics cannot readily be converted into database records, making automated literature extraction an important direction in materials informatics \citep{ref024,ref026,ref034}. Advances in large language models (LLMs) offer new technical approaches to schema-constrained extraction from heterogeneous scientific texts; their potential and limitations have been preliminarily examined in studies of materials and chemical data extraction \citep{ref008,ref011,ref040}.

Early research on materials text mining established foundational tools and resources for chemical information recognition, relation extraction, and terminology normalization, including ChemDataExtractor, databases of magnetic transition temperatures, and named entity recognition pipelines for materials \citep{ref006,ref044,ref050}. Research on word embeddings, inorganic synthesis recipe datasets, the SOFC-Exp corpus, and battery materials databases further demonstrated different approaches to representing and structuring literature-derived knowledge \citep{ref012,ref019,ref027,ref046}. Building on this foundation, MatSciBERT and automatically generated corpora strengthened the connection between domain language modeling and information extraction \citep{ref018,ref051}. Studies addressing superalloys, photovoltaic devices, polymers, integrated text and table extraction, and stress--strain properties further extended automated pipelines to diverse materials and parameter types \citep{ref002,ref028,ref041,ref049,ref054}.

Recent work has extended the use of LLMs to polymer property extraction, question answering over perovskite literature, and materials language processing \citep{ref004,ref017,ref042}, while also investigating retrieval and tool augmentation, the construction of battery-recipe knowledge bases, and multimodal information extraction for nanomaterials \citep{ref029,ref033,ref038}. Meanwhile, studies such as MatSci-NLP, PcMSP, materials named entity recognition, and MaScQA have expanded evaluation to language tasks, synthesis action graphs, entity extraction, and domain-specific knowledge question answering, respectively \citep{ref020,ref043,ref052,ref053}. Document-level parameter extraction, n-ary property relation modeling, and information recovery from tables further highlight the importance of experimental conditions, scientific roles, and evidence organization for structured output (\citealp{ref015}; \citealp{ref035}; \citealp{ref036}; \citealp{ref037}). Collectively, these studies establish a foundation for automated processing of materials literature, although their task definitions, input conditions, and evaluation units differ.

Document-level parameter extraction across heterogeneous materials systems still faces three challenges. First, many existing pipelines are designed around specific materials or parameter schemas, making their results difficult to compare directly across domains \citep{ref007,ref017,ref029,ref042}. Second, material variants, experimental groups, substrates, and measurement conditions within a paper may share similar expressions. Correct field-value recovery therefore depends on instance-level attribution and cross-paragraph evidence integration; multidimensional correspondences in tables further increase the difficulty (\citealp{ref005}; \citealp{ref015}; \citealp{ref035}; \citealp{ref036}). Third, the accessible evidence in retrieval-augmented pipelines depends on the retrieval scope, knowledge-base retention state, and task prompts. These conditions must be specified explicitly in the evaluation protocol \citep{ref031,ref038}. Accordingly, systematic comparisons across materials systems, models, and extraction configurations require unified instance-alignment and field-counting rules.

To address these issues, this study introduces MatExtractBench, an entity-conditioned document-level parameter extraction benchmark for materials science literature. The task takes a predefined set of target material instances and a parameter schema as input, requiring models to extract the corresponding parameters from paper-based evidence and correctly associate field values with their target instances. The benchmark covers nanozymes and PSCs, two materials systems with distinct parameter structures, including composition, morphology, experimental conditions, catalytic and photovoltaic performance, and fabrication-process variables. At the data level, it applies unified principles for evidence verification, field normalization, and instance organization. At the evaluation level, RagFlow is used for paper-level evidence ingestion, entity-conditioned retrieval, structured extraction, and result saving, followed by removal of the corresponding content from the knowledge base. Instance alignment and field scoring are conducted on the saved predictions. A single PDF constitutes the processing unit for nanozymes, whereas the main article and its supplementary file constitute one evidence group for PSCs.

In addition to this main workflow, three control configurations are established: direct extraction without predefined field names, retrieval with a cumulative knowledge base, and direct extraction conditioned on the paper and material identities. Fifteen models are evaluated in both domains. The configurations share the evaluation targets and counting protocol, while their input constraints and evidence organization are reported separately. The comparisons characterize the relationship between complete extraction configurations and final field recovery quality, as well as performance differences across the two materials systems.

The main contributions of this study are as follows:

\begin{itemize}
\item \textbf{A cross-domain parameter extraction benchmark.} Two domain-specific datasets, covering nanozymes and PSCs, are developed to evaluate document-level parameter extraction conditioned on predefined target instances.
\item \textbf{Consistent annotation and evaluation units.} Material instances and heterogeneous parameters are organized using literature evidence, with unified field semantics, missing-value encoding, instance alignment, and value-level counting rules.
\item \textbf{An automated paper-level isolated workflow.} A single PDF for each nanozyme paper and the main--supplementary PDF group for each PSC paper are treated as separate processing units; the associated knowledge-base content is cleared after extraction and result saving.
\item \textbf{Cross-model and cross-configuration comparisons.} Thirteen locally deployed models and two API models are evaluated in four extraction configurations to examine performance differences associated with evidence management, input conditions, and materials parameter structures.
\end{itemize}

\begin{figure}[tp]
\centering
\includegraphics[
  width=\linewidth,
  height=.62\textheight,
  keepaspectratio
]{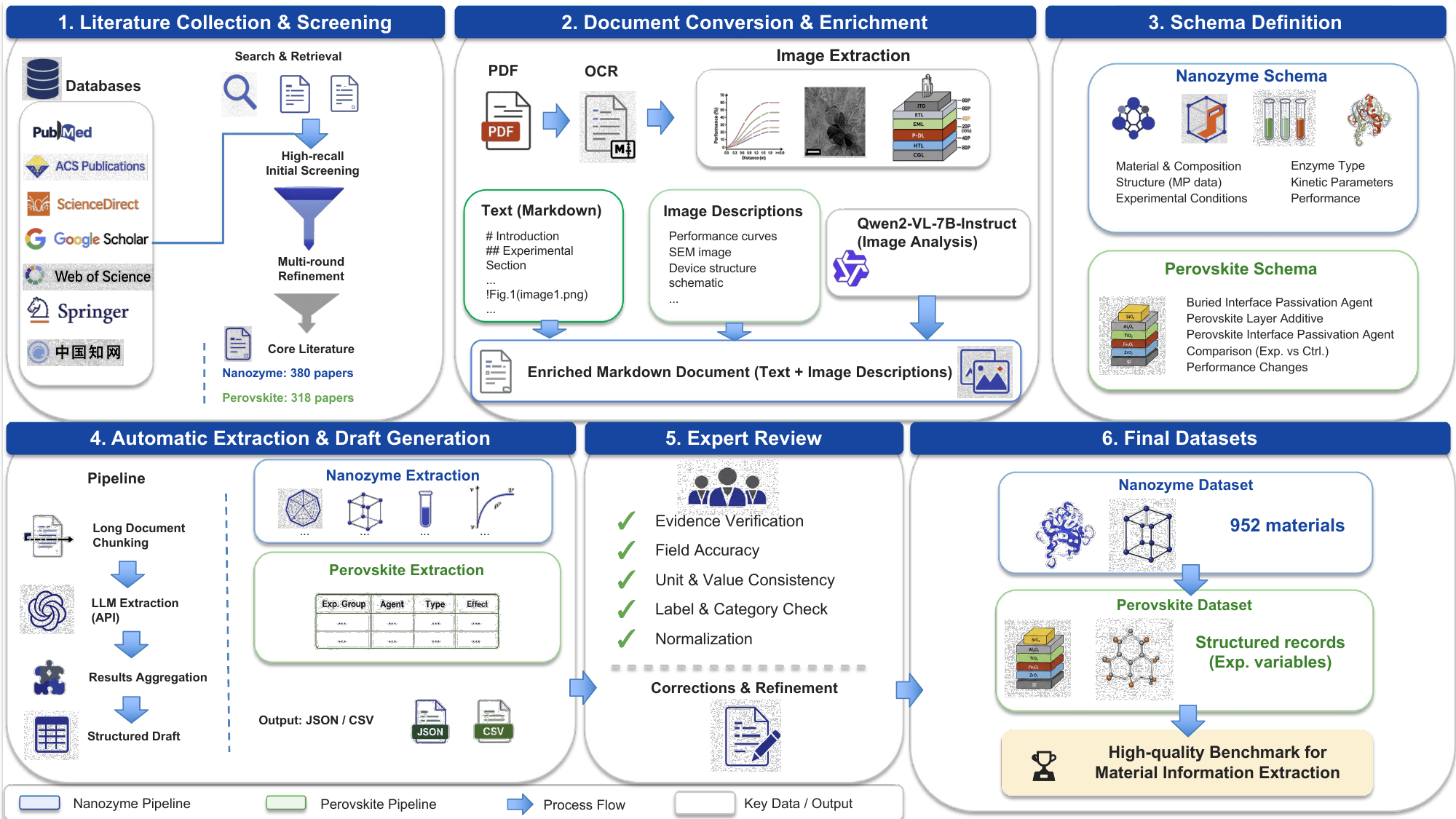}
\caption{Literature screening, assisted annotation, and expert verification workflow for the nanozyme and PSC datasets.}
\label{fig:1}
\end{figure}
\section{Methods}
\subsection{Domain-specific dataset construction}
The MatExtractBench data construction process consists of literature collection and screening, document parsing, parameter-schema definition, LLM-assisted annotation, expert verification, and structured data organization. Public data resources supplement materials information, whereas evidence in the source documents determines the parameter labels eligible for evaluation. The nanozyme and PSC datasets follow the same data-organization principles while retaining differences in field semantics, experimental conditions, and evidence distribution across the two materials systems.

\textbf{Literature collection and screening.} Source papers were collected through indexing and publishing platforms, including Web of Science, PubMed, Google Scholar, ACS, ScienceDirect, and Springer. Domain-specific search terms were compiled separately for nanozymes and PSCs. Keyword-frequency and topical-relevance analyses were used for initial screening, followed by manual verification using titles, abstracts, and full texts.

The initial nanozyme candidate pool contained 1,742 papers, of which 380 were retained after multistage screening. Selection focused on topical relevance, completeness of experimental information, and the availability of evidence needed for parameter extraction. The PSC corpus followed the same search and verification principles, with particular emphasis on studies reporting the introduction methods and device-layer locations of passivating agents or additives, experimental-versus-control-group differences, and associated performance outcomes, thereby supporting fine-grained extraction of process-related variables.

\textbf{Material-level data organization.} Previous nanozyme research has combined literature-derived data with externally sourced structural properties for data-driven materials design \citep{ref025}. This study constructs a nanozyme dataset containing 952 material-level structured records by integrating public data resources with experimental information from the original literature, distinguishing evaluation labels from auxiliary attributes according to their evidentiary sources. Nonempty parameter values verified against the corresponding source documents are included in the primary evaluation; properties supplemented from external resources such as the Materials Project, but unsupported by the corresponding documents, are retained only as auxiliary information \citep{ref022}. Material composition, morphology, experimental conditions, and catalytic performance are organized using consistent field definitions so that the evaluated values correspond to evidence available to the model in the source literature.

The PSC dataset is organized around experimental variables introduced during device fabrication, linking chemical composition, application methods, device-layer locations, and photovoltaic performance. During dataset construction, experimental and control-group information from the main text and supplementary materials is considered jointly. Variables are classified as buried-interface passivating agents, perovskite-interface passivating agents, or perovskite-layer additives according to their principal introduction method and functional location. Parameters such as chemical formulae, composition ratios, amounts, and units are also recorded. The mutual exclusivity of these categories applies to specific experimental-use records rather than to the chemical compounds themselves. The final records follow the predefined schema and include composition-related attributes, antisolvents, passivating agents and additives, and key photovoltaic metrics.

\textbf{Document preprocessing.} Document parsing and optical character recognition (OCR) tools are used to convert PDFs into Markdown text \citep{ref003,ref047}, retaining the sectional and paragraph structure of the main articles and supplementary materials to support subsequent chunking and evidence localization. A locally deployed Qwen2-VL-7B-Instruct model is also used to generate auxiliary descriptions of scientific figures \citep{ref048}, supplementing visual information about material morphologies, device structures, experimental schematics, and performance trends.

PSC preparation conditions, interface treatments, and device performance are often reported at different locations in the main text and supplementary materials. A unified machine-readable representation facilitates the joint organization of this evidence; however, automated parsing and image descriptions may still omit information or introduce transcription errors, requiring parameter annotations to be checked against the original documents.

\textbf{Parameter-schema definition.} Domain experts separately define the parameter extraction schemas for nanozymes and PSCs. The nanozyme schema covers material names, chemical composition, physical properties, experimental conditions, enzyme-like activity types, and kinetic parameters. The PSC schema covers chemical representations of fabrication variables, their introduction methods and functional locations, and photovoltaic performance. Target instances and their annotated records are determined from the original papers, establishing correspondences among instance identifiers, experimental conditions, and field values. Variables that fall into multiple process categories are classified according to their principal introduction method and location in each specific experimental-use record.

Gold-standard field encoding follows a unified semantic convention: parameters not reported in the paper or lacking sufficient evidence are assigned JSON null; for binary fields such as elemental presence, 1 and 0 indicate evidence-supported presence and absence, respectively, while indeterminate cases remain null. Numerically meaningful zero values are retained for numeric fields. The absence of any mention of an element in a paper does not, by itself, constitute evidence that the element is absent. These reference-label conventions are applied during candidate annotation and expert verification. They are distinguished from the nanozyme prompt's binary-only output constraint, as described below.

\textbf{LLM-assisted annotation.} OCR-converted Markdown documents are split into chunks and passed through a DeepSeek-R1 API under a parameter schema and system-prompt constraints to generate candidate records \citep{ref009}. Nanozyme annotations cover composition, morphology, physical properties, enzyme-like activity, and kinetic parameters. PSC annotations cover the chemical information, process roles, device-layer locations, and performance metrics of passivating agents and additives. Chunk-level results are consolidated into document-level JSON records for subsequent expert verification and correction. The API model used for draft annotation is recorded separately from the evaluated models, and its candidate outputs are not directly used as gold-standard labels.

\textbf{Expert review and final dataset construction.} Domain experts verify candidate records field by field against the original articles, focusing on evidence supporting each parameter value, instance-to-field correspondence, the attribution of experimental conditions and process roles, and consistency of numerical values, units, and chemical representations. Extraction errors, instance mismatches, and semantic ambiguities are corrected using the main text, tables, and supplementary materials. The final gold standard is determined from documentary evidence and manual verification; externally supplemented values lacking documentary support are not treated as valid extraction labels.

Following verification and normalization, each evaluation sample consists of a source-paper evidence group, a predefined set of target instances, and the corresponding gold-standard records. Each target instance in a paper is associated with its own set of parameter fields; unreported or indeterminate fields are retained under the unified missing-value convention. This organization provides a consistent data basis for subsequent instance-level matching and field-level counting.

\textbf{Dataset characteristics.} Table 1 compares the record units, representative fields, and extraction requirements of the two datasets. Both follow the same evaluation principles, while their parameter types and evidence-association structures are complementary.

The nanozyme dataset is organized by material instance, with fields describing composition, morphology, particle size, surface treatment, dispersion medium, reaction conditions, substrate, and enzyme kinetic parameters. Composition and morphology may be supported by material descriptions or tables, whereas parameters such as pH, temperature, substrate, Km, Vmax, and Kcat depend on specific testing conditions. Extracting the latter requires distinguishing material variants, substrates, and reaction settings within the same paper to avoid merging values obtained under different experimental conditions into a single record.

\begin{table}[t]
\centering
\caption{Parameter schemas and evidence organization characteristics of the nanozyme and PSC datasets.}
\label{tab:1}
\footnotesize
\renewcommand{\arraystretch}{1.12}
\setlength{\tabcolsep}{4pt}
\begin{tabularx}{\linewidth}{@{}
>{\raggedright\arraybackslash}p{.13\linewidth}
>{\raggedright\arraybackslash}p{.17\linewidth}
>{\raggedright\arraybackslash}X
>{\raggedright\arraybackslash}X@{}}
\toprule
Dataset & Record unit & Representative schema fields & Extraction characteristics\\
\midrule
Nanozyme dataset & Material-level nanozyme instance & Nanozyme name; classification; elemental indicators for O, N, P, S, Si, Se, B, F, Cl, Br, and I; primary metal oxidation state; shape; size; surface treatment; dispersion medium; pH; temperature; substrate; Km; Vmax; Kcat & Covers material composition, morphology, reaction conditions, and enzyme kinetic parameters; condition-dependent fields must be linked to their respective substrates and experimental settings.\\
PSC dataset & Target experimental or device record & PCE; VOC; JSC; FF; buried-interface passivating agents; perovskite-layer additives; perovskite-interface passivating agents & Extracts chemical constituents for specified targets, determines fabrication stages and device-layer locations, and associates the corresponding photovoltaic metrics.\\
\bottomrule
\end{tabularx}
\end{table}
The PSC dataset is organized around fabrication variables and device performance, covering power conversion efficiency (PCE), open-circuit voltage (VOC), short-circuit current density (JSC), fill factor (FF), and fields for buried-interface passivating agents, perovskite-layer additives, and perovskite-interface passivating agents. Given known target instances, the model must extract the relevant chemical constituents from experimental descriptions and determine their process roles using the fabrication stage, device-layer structure, and experimental--control group relationships. This task focuses on parameter extraction and role attribution for specified instances. Its differences from the nanozyme task lie primarily in field semantics, conditional dependencies, and evidence distribution; cross-domain results must be interpreted in light of these factors.

\begin{figure}[tp]
\centering
\includegraphics[
  width=\linewidth,
  height=.62\textheight,
  keepaspectratio
]{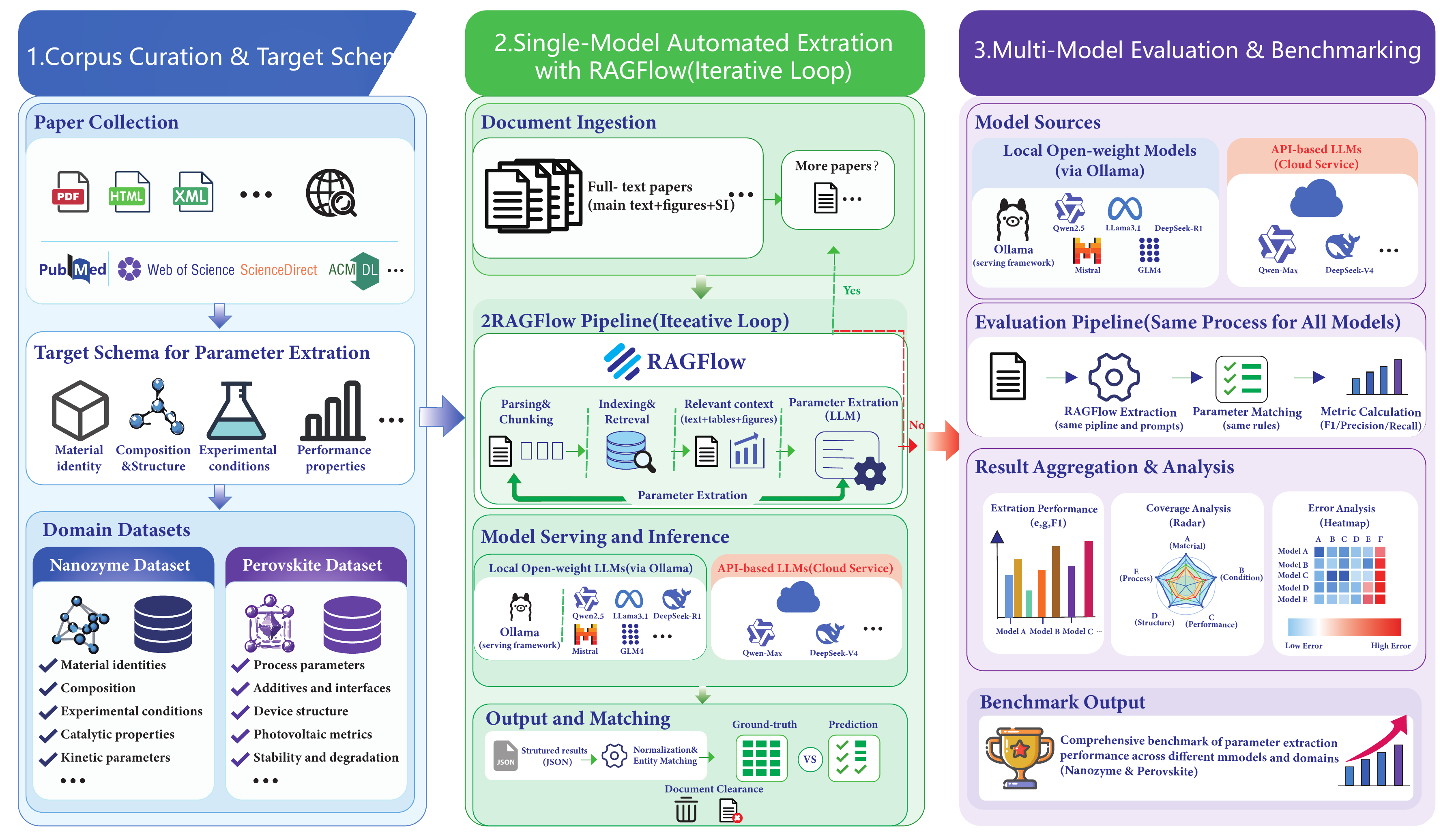}
\caption{Paper-level isolated evaluation framework of MatExtractBench.}
\label{fig:2}
\end{figure}
\subsection{RagFlow-based automated evaluation workflow}
To evaluate parameter recovery for specified material instances, a paper-level isolated workflow is implemented using RagFlow. Each paper undergoes evidence ingestion, parsing and chunking, entity-conditioned retrieval, structured extraction, and result saving, after which the corresponding contents are deleted from the knowledge base. Instance alignment and field scoring are performed on the saved prediction records and do not require the original files to remain in the knowledge base. A single PDF serves as the evidence unit for nanozymes, whereas the main PDF and corresponding supplementary PDF jointly constitute an evidence unit for PSCs. Thus, the paper defines the retrieval scope, the target instance determines record attribution, and the instance--field combination forms the basic counting unit.

Let the benchmark corpus be denoted by:

\begin{equation}
D = \{d_i\}_{i=1}^{N}
\label{eq:1}
\end{equation}
where $d_i$ denotes the logical evidence unit associated with the i-th paper and N is the total number of papers. Let $G_i$ denote the set of source files for that paper: $G_i$ = \{$p_i$\} for the nanozyme task and $G_i$ = \{$p_i$, $a_i$\} for the PSC task, where $p_i$ and $a_i$ are the main PDF and its corresponding supplementary PDF. Both files from the same paper provide evidence jointly and are not counted as two independent papers. For each paper $d_i$, define the target-instance set:

\begin{equation}
M_i = \{m_{i,k}\}_{k=1}^{K_i}
\label{eq:2}
\end{equation}
where $M_i$ is the set of target material instances predefined by benchmark annotations; $m_{i,k}$ denotes its k-th target instance, and $K_i$ is the number of target instances in that paper. In the main workflow, both PSC and nanozyme targets are supplied individually through the \texttt{\{name\}} placeholder in successive extraction requests. The model extracts parameters for these known instances. The corresponding gold-standard annotations are defined as:

\begin{equation}
Y = \{y_{i,k} \mid d_i \in D,\ m_{i,k} \in M_i\}
\label{eq:3}
\end{equation}
Let S be the predefined field schema and $f_{\theta}$ the evaluated model. Given ($d_i$, $m_{i,k}$, S), the task is to generate a structured prediction $\hat{y}_{i,k}$ for target instance $m_{i,k}$ and compare it with the instance's gold-standard record $y_{i,k}$. When $K_i$ > 1, the targets correspond to separate parameter records and field values must not be merged across instances. Because target identifiers are provided as input, the reported metrics do not independently evaluate open-ended material entity discovery or named entity recognition.

\textbf{Paper-level evidence ingestion and isolation.} During each processing cycle, the source-file set $G_i$ is imported into RagFlow, parsed, and chunked to construct the joint retrieval set:

\begin{equation}
C_i = \bigcup_{g\in G_i}\operatorname{Chunk}(g)
\label{eq:4}
\end{equation}
where $C_i$ is the union of text chunks from all source files associated with the current paper. The PSC main text and supplementary file are both included in retrieval, enabling preparation conditions, interface treatments, and photovoltaic performance reported in separate files to be associated within the same paper-level evidence scope. The main workflow's retrieval set contains no previously processed papers.

\textbf{Entity-conditioned retrieval and extraction.} Relevant evidence is retrieved from $C_i$ according to the field schema S and target-instance identifier $m_{i,k}$:

\begin{equation}
R(d_i,m_{i,k}) = \operatorname{Retrieve}(C_i,S,m_{i,k})
\label{eq:5}
\end{equation}
where R($d_i$, $m_{i,k}$) is the evidence set pertaining to the target instance and fields within the current paper. PSC retrieval results may include chunks from both the main article and its supplement, constituting cross-file evidence integration within a single paper. This scope restriction applies to the main workflow; the cumulative-knowledge-base control is described separately below.

The extraction procedure for the evaluated model is summarized at the target-record level as:

\begin{equation}
\hat{y}_{i,k} = f_{\theta}\!\left(m_{i,k},S,R(d_i,m_{i,k})\right)
\label{eq:6}
\end{equation}
Here, $\hat{y}_{i,k}$ is the assembled parameter prediction for target instance $m_{i,k}$. Equation~\ref{eq:6} summarizes the final target-level record: for PSCs, this record is the single data row parsed from a CSV response for one specified target; for nanozymes, it is assembled from three sequential field-group responses for the same target. The number of model queries is therefore specified by the domain-specific prompting procedure.

\textbf{Task-specific prompting and output assembly.} Figure~\ref{fig:task-specific-prompts-psc}(a) shows the PSC template. The placeholders \texttt{\{doc\_id\}} and \texttt{\{name\}} identify the source paper and one predefined target instance, while \texttt{\{document\_text\}} contains the paper-scoped text evidence supplied to the extraction step. In the main retrieval workflow, this evidence may come from both the main article and its corresponding supplement. Each request specifies one target and returns CSV containing a header and exactly one data row, preserving \texttt{\{name\}} in the \texttt{catalyst} column. The seven extracted parameter fields are PCE, VOC, JSC, FF, buried-interface passivating agents, perovskite-layer additives, and perovskite-interface passivating agents. Together with the target name, these form an eight-column record. Electrical values are returned without units; chemical entries use formulas, and fields containing comma-separated values are enclosed in double quotation marks. For papers with multiple predefined targets, this request is repeated separately for each target before paper-level cleanup.

Figure~\ref{fig:task-specific-prompts-nanozyme}(b) shows the shared nanozyme instructions and the three field-group templates. Each query specifies one target through \texttt{\{name\}}, its source through \texttt{\{doc\_id\}}, and the supplied evidence excerpts through \texttt{\{evidence\}}. For each predefined nanozyme, the model is queried sequentially for three groups: (1) classification and the 11 elemental indicators O, N, P, S, Si, Se, B, F, Cl, Br, and I; (2) metal oxidation state, shape, size, and surface treatment; and (3) dispersion medium, optimal pH, optimal temperature, substrate, Km, Vmax, and Kcat. Each call includes the shared instructions and only its corresponding field template. The response is a single line of labeled \texttt{field=value} assignments separated by \texttt{|||}, with no header, JSON, Markdown, or explanation; each extracted value must contain fewer than 12 words. The three responses are parsed by their field labels and combined into 23 extracted parameter fields. The predefined target name is added as \texttt{Nanozyme Name}, giving a 24-column record. This assembly combines complementary field groups belonging to the same target; it does not combine predictions attributed to different targets.

\textbf{Multi-instance result alignment.} Predictions are first aligned with gold-standard records using a unique instance identifier or a normalized material name that maps unambiguously to the target, rather than by row order or an optimal pairing based on the similarity of evaluated parameter values. Experimental variants sharing a material name are matched separately only when disambiguating identifiers are provided in the input. Records whose attribution cannot be uniquely determined do not count as valid field matches, nor are they combined across rows to construct a correct record. Multicomponent fields are compared as complete fields following predefined multivalue normalization rules; they are not split into new target instances. The counting rules for unmatched and extra records are specified below.

\textbf{Field normalization.} The two prompts use different response formats, but both are converted into target-indexed field records before comparison. A literal \texttt{null} in a PSC CSV cell or a nanozyme non-element assignment is normalized to a missing value; raw model responses are not required to be JSON. Reported nanozyme classifications use CAT, HYL, OXD, POD, or SOD. The nanozyme prompt constrains the 11 elemental predictions to 0 or 1 and excludes elements appearing only in assay reagents. This output constraint does not establish evidence for elemental absence when the source is indeterminate. Gold-standard indeterminate elemental fields retain null under the annotation convention above; a binary prediction for such a field remains nonempty and is scored using the rules below.

To reduce comparison discrepancies arising from equivalent expressions, the same normalization operator N(·) is applied to predictions and gold-standard records:

\begin{equation}
\tilde{y}_{i,k} = N(\hat{y}_{i,k})
\label{eq:7}
\end{equation}
\begin{equation}
\tilde{y}_{i,k}^{*} = N(y_{i,k})
\label{eq:8}
\end{equation}
where $\tilde{y}_{i,k}$ and $\tilde{y}^{*}_{i,k}$ denote the normalized prediction and normalized gold-standard record, respectively. Instance identifiers and name aliases are used for row alignment, while composition expressions, numerical formats, and units are transformed according to predefined field-specific rules. Only expressions unambiguously determined to be equivalent by these rules are treated as matches.

For any schema field $s_j$ $\in$ S, the field-level correctness indicator is defined as:

\begin{equation}
c_{i,k,j} = \mathbf{1}\!\left[
\tilde{y}_{i,k,j} = \tilde{y}_{i,k,j}^{*}
\right]
\label{eq:9}
\end{equation}
where $\mathbf{1}[\cdot]$ denotes the indicator function, which equals 1 when the normalized predicted and gold-standard field values match and 0 otherwise. This indicator characterizes value-level agreement; the subsequent performance counts also depend on whether the corresponding fields are empty.

\textbf{Evaluation metrics.} This study uses a value-level counting protocol based on instance alignment, with the triplet of paper, target instance, and parameter field (i, k, j) as the basic evaluation unit. After row matching and field normalization, outcomes are counted as follows: one true positive (TP) if both the gold-standard and predicted values are nonempty and match; one false negative (FN) if the gold-standard value is nonempty and the prediction is empty; one false positive (FP) if the gold-standard value is empty and the prediction is nonempty; one FP and one FN if both are nonempty but disagree, reflecting the generation of an incorrect predicted value and the failure to recover the correct reference value; and one true negative (TN) if both are empty. Missingness is determined after output parsing and normalization, using the semantics of JSON null in the structured records. Numerically meaningful zero values and binary predictions of 0 are treated as nonempty values. Thus, a predicted elemental value of 0 or 1 against an indeterminate null reference contributes an FP under the same empty-reference rule.

For target instances without a valid matched prediction row, nonempty gold-standard fields contribute to FN. Nonempty evaluated fields in duplicate or ambiguous records that cannot be uniquely attributed, and in additional predictions outside the target set, contribute to FP; fields are not concatenated across instances. Multivalue fields are evaluated as complete normalized field values without component-level partial credit. All extraction configurations use the same target set, field set, and counting rules.

Precision, recall, and F1 are computed from these counts to quantify extraction performance:

\begin{equation}
\mathrm{Precision} = \frac{\mathrm{TP}}{\mathrm{TP}+\mathrm{FP}}
\label{eq:10}
\end{equation}
\begin{equation}
\mathrm{Recall} = \frac{\mathrm{TP}}{\mathrm{TP}+\mathrm{FN}}
\label{eq:11}
\end{equation}
\begin{equation}
F_1 = \frac{2\times\mathrm{Precision}\times\mathrm{Recall}}{\mathrm{Precision}+\mathrm{Recall}}
\label{eq:12}
\end{equation}
Precision is the proportion of nonempty predictions that match correctly, recall is the proportion of nonempty gold-standard values successfully recovered, and F1 is their harmonic mean. TN is recorded separately but does not contribute to these three metrics. For each model, dataset, and configuration, TP, FP, and FN are first aggregated across all instance--field units and extra prediction records, after which the corresponding metrics are calculated. Cross-model averages are used only for descriptive comparisons. Any metric with a zero denominator is assigned a value of 0.

\textbf{Result saving and knowledge-base cleanup.} After the individual PSC CSV responses or all three nanozyme field-group outputs have been parsed and assembled for every target in the current paper, both the raw responses and assembled records are saved with their paper and target identifiers. The knowledge base is then cleared. Subsequent instance matching, field normalization, and scoring are performed using the saved records. The cleanup operation is written as:

\begin{equation}
K \leftarrow K\setminus G_i
\label{eq:13}
\end{equation}
where K denotes the set of documents in the knowledge base. For the nanozyme task, the current PDF and its associated retrieval content are removed; for the PSC task, the main PDF, its corresponding supplementary PDF, and their associated retrieval content are removed together. The next paper is imported only after this evidence group has been fully cleared, maintaining paper-level evidence boundaries throughout sequential processing.

\textbf{Multi-model evaluation.} Each evaluated model sequentially processes the papers in the same task, and field counts are aggregated separately for each model. Instance-alignment and scoring rules are consistent within a configuration, while differences in evidence organization, input constraints, and knowledge-base management are explicitly identified as experimental conditions across configurations.

The workflow integrates evidence-scope management, instance-level parameter attribution, and field evaluation into a unified protocol for comparing the quality of structured parameter recovery across models under specified task conditions.

\begin{figure}[tp]
\centering
\begin{promptbox}
Perovskite solar-cell parameter extraction prompt

Target perovskite: {name}
Source document: {doc_id}

Task:
Extract the specified parameters for {name} using only
the evidence provided in the current document.

Parameters to extract:

1. "PCE(%)":
   Power conversion efficiency (%), numerical value.
2. "VOC(V)":
   Open-circuit voltage (V), numerical value.
3. "JSC(mA cm-2)":
   Short-circuit current density (mA/cm2),
   numerical value.
4. "FF(%)":
   Fill factor (%), numerical value.
5. "buried interface passivation agent":
   Chemical formula(s) of passivation agents
   applied before the perovskite layer
   (on the transport layer or in the
   SnO2 solution).
   Multiple values: comma-separated.
6. "perovskite layer additives":
   Chemical formula(s) of additives dissolved
   in the perovskite precursor solution.
   Multiple values: comma-separated.
7. "perovskite interface passivation agent":
   Chemical formula(s) of passivation agents
   applied after perovskite annealing.
   Multiple values: comma-separated.

Extraction rules:
- Use only information from the current document.
- Extract exactly one record for {name}.
- Use {name} exactly as the target identifier.
- Associate every parameter value with {name}.
- Do not mix parameter values from different
  materials, device variants, experimental
  groups, or fabrication conditions.
- If a parameter is not reported, use null.
- For passivation agents and additives,
  output only chemical formulas.
- For electrical parameters, output
  numerical values without units.
- Separate multiple chemical formulas
  using commas.
- Enclose CSV fields containing commas
  in double quotation marks.
- Do not return records for other targets.

Output format (CSV):
Return the header followed by exactly one data row.
catalyst,PCE (%),VOC (V),JSC (mA cm-2),FF (%),buried interface passivation agent,perovskite layer additives,perovskite interface passivation agent

Output example (illustrative values):
{name},21.2,1.12,23.5,80.1,KBr,MACl,"PEAI, PMMA"

Please extract parameters for {name}
from the following text:

{document_text}
\end{promptbox}

\caption{Task-specific extraction prompts in the
RAGFlow-based evaluation pipeline:
(a) single-target perovskite solar cell
parameter extraction.}
\label{fig:task-specific-prompts-psc}
\end{figure}


\begin{figure}[tp]
\ContinuedFloat
\captionsetup{list=off}
\centering
\begin{promptbox}
Nanozyme parameter extraction prompt

Target nanozyme: {name}
Source document: {doc_id}

Source evidence:
{evidence}

Task:
Extract parameters for the specified nanozyme using only
the evidence excerpts from the selected document.

For {name}, return labeled field=value assignments
on one line separated by |||.

Every value must retain its exact field label.
Do not include a header, JSON, Markdown, or explanation.
Keep each extracted value under 12 words.

Extraction rules:
- Use only evidence from the selected document.
- Extract parameters for the predefined target nanozyme.
- Do not mix values from different materials, variants,
  substrates, or experimental conditions.
- For non-element fields, use null when the
  parameter is not reported.
- Classification must use CAT, HYL, OXD, POD, or SOD.
- For O, N, P, S, Si, Se, B, F, Cl, Br, and I,
  output 1 if the target nanozyme itself contains
  the element and 0 if it does not.
- Never output null for elemental composition fields.
- Exclude elements appearing only in substrates,
  buffers, solvents, or other assay reagents.

Use the exact field template for the current group:

Group 1: Composition and classification

Classification=<value>|||O=<value>|||N=<value>
|||P=<value>|||S=<value>|||Si=<value>
|||Se=<value>|||B=<value>|||F=<value>
|||Cl=<value>|||Br=<value>|||I=<value>

Group 2: Metal oxidation state and physical properties

Metal oxidation state=<value>|||Shape=<value>
|||Size (nm)=<value>|||Surface treatment=<value>

Group 3: Experimental conditions and kinetic parameters

Dispersion medium=<value>|||Optimal pH=<value>
|||Optimal temperature (C)=<value>
|||Substrate=<value>|||Km (mM)=<value>
|||Vmax (uM s-1)=<value>|||Kcat (s-1)=<value>

Execution:
For each predefined nanozyme, the model is queried
sequentially using the three field-specific templates.
Only the corresponding field template is supplied
in each model call.

The three group-level outputs are consolidated into
one structured record containing 24 columns,
including Nanozyme Name.
\end{promptbox}

\caption[]{Task-specific extraction prompts in the
RAGFlow-based evaluation pipeline:
(b) nanozyme extraction using shared instructions
and three sequential field-group queries.}
\label{fig:task-specific-prompts-nanozyme}
\end{figure}

\subsection{Benchmark formalization and experimental configurations}
MatExtractBench consists of a corpus of scientific papers, gold-standard records, a field schema, and evaluation protocols. The primary task is entity-conditioned document-level parameter extraction. Additional direct-extraction and cumulative-knowledge-base configurations are used to compare the overall performance of workflows with different input conditions and evidence-management strategies.

Formally, MatExtractBench is defined as:

\begin{equation}
B = (D,Y,S,P,M)
\label{eq:14}
\end{equation}
where D denotes the paper corpus, Y the gold-standard records, S the field schema, P the set of evaluation protocols, and M the model-level and field-level statistics. Different protocols share the same evaluation targets but separately specify how evidence is organized and what inputs are available to the models.

The paper serves as the evidence-organization unit, whereas each paper--predefined-target-instance pair is the parameter-record unit. For target instance $m_{i,k}$ in paper $d_i$, the evaluation instance is defined as:

\begin{equation}
b_{i,k} = (d_i,m_{i,k},y_{i,k},S),\qquad m_{i,k}\in M_i
\label{eq:15}
\end{equation}
A single paper may provide multiple evaluation instances, each associated with a set of gold-standard fields. The main text, tables, and corresponding supplement may provide evidence for the same target, but predictions must retain their instance-level attribution. Target identifiers are supplied in the main workflow; the input conditions of the other configurations are given in Table 2.

The two datasets use the same instance-level and field-level evaluation units. The nanozyme task examines composition, morphology, experimental conditions, and enzymatic parameters. The PSC task examines fabrication variables, their device-layer roles, and photovoltaic performance for specified material or experimental records. The names of passivating agents and additives in the PSC schema are values to be extracted into fields; models are not asked to determine the set of evaluation targets independently.

The normalized document-level gold standard retains all target instances and their field assignments:

\begin{equation}
\tilde{y}_{i}^{*} = \{N(y_{i,k})\}_{k=1}^{K_i}
\label{eq:16}
\end{equation}
Each field value retains the index of its target instance, while numerical values, units, and chemical expressions are transformed according to predefined rules.

For each model and extraction configuration, field-comparison results are aggregated across papers to calculate task-level precision, recall, and F1; an entire paper is not classified as a single correct or incorrect prediction. The corpus-level evaluation is expressed as:

\begin{equation}
\begin{aligned}
E_{\theta}^{B} = \operatorname{Metric}\!\Biggl(&
    \sum_i\sum_k\mathrm{TP}_{i,k},\\
&   \sum_i\sum_k\mathrm{FP}_{i,k}+\sum_i\mathrm{FP}_{i,\mathrm{extra}},\\
&   \sum_i\sum_k\mathrm{FN}_{i,k}\Biggr).
\end{aligned}
\label{eq:17}
\end{equation}
Here, $\mathrm{TP}_{i,k}$, $\mathrm{FP}_{i,k}$, and $\mathrm{FN}_{i,k}$ are instance-level field counts, while $\mathrm{FP}_{i,\mathrm{extra}}$ counts nonempty predicted fields in additional records that cannot be matched to a target instance. Metric denotes calculation using Eqs. (10)--(12). The configurations are evaluated on the same targets, but their input conditions are not identical; therefore, the comparisons reflect the aggregate performance of each complete configuration.

Cross-model means for each configuration are calculated as the equally weighted arithmetic mean of the 15 individual model metrics. They summarize the evaluated model set and are not equivalent to corpus-level metrics recomputed after pooling predictions from different models.

\textbf{Experimental configurations.} Fifteen models are compared across two datasets under four extraction configurations. RAG-I denotes the main paper-level isolated workflow; Direct-U, RAG-A, and Direct-C denote direct extraction without predefined field names, cumulative-knowledge-base retrieval, and direct extraction conditioned on paper and material identities, respectively. Figure~\ref{fig:task-specific-prompts-psc} documents the task-specific prompts of the main workflow. Table 2 specifies the input conditions and knowledge-base management strategies of the four configurations; the main-workflow templates do not imply identical prompting conditions for all controls.

\begin{table}[t]
\centering
\caption{Evidence organization, input conditions, and knowledge-base management across four extraction configurations.}
\label{tab:2}
\footnotesize
\renewcommand{\arraystretch}{1.12}
\setlength{\tabcolsep}{4pt}
\begin{tabularx}{\linewidth}{@{}
>{\raggedright\arraybackslash}p{.23\linewidth}
>{\raggedright\arraybackslash}X
>{\raggedright\arraybackslash}p{.25\linewidth}@{}}
\toprule
Extraction configuration & Evidence acquisition and input conditions & Knowledge-base management\\
\midrule
RAG-I
Paper-level isolated retrieval & Retrieves evidence using target material instances and the parameter schema of the current paper; nanozymes use one PDF and PSCs jointly retrieve the main text and supplement. & After all targets in the current paper have been processed, deletes the entire file group and associated retrieval content.\\
Direct-U
Direct extraction without predefined fields & Directly invokes the model to process the paper without RagFlow retrieval; fixed parameter names are not supplied in the prompt. & Does not use a RagFlow knowledge base.\\
RAG-A
Cumulative-knowledge-base retrieval & Uses RagFlow for retrieval and extraction; previously processed papers remain in the knowledge base. & Disables automatic cleanup after processing each paper, allowing the knowledge base to accumulate.\\
Direct-C
Paper- and material-conditioned direct extraction & Does not use RagFlow retrieval; the current paper ID and material name specify the extraction target. & Does not use a RagFlow knowledge base.\\
\bottomrule
\end{tabularx}
\end{table}
\textbf{Model deployment and invocation.} The main benchmark comparison comprises 13 models deployed locally via Ollama and two models accessed through service APIs. DeepSeek-V4 and Qwen-Max are accessed through their respective APIs in all four configurations and are not local Ollama models. The remaining models are Qwen2.5-7B/14B, Llama3-8B, Llama3.1-8B, DeepSeek-R1-7B/14B/32B, Mistral-7B, Mistral-Nemo-12B, GLM4-9B, Granite3.3-8B, Gemma-7B, and Gemma2-9B. Model identifiers used in the experiments are retained in the results tables. The distinction between local deployment and API access describes the execution mode and does not imply whether model weights are openly available.

The technical sources for the local models include reports on the Qwen2.5, Llama 3, DeepSeek-R1, Mistral 7B, and GLM families \citep{ref009,ref016,ref023,ref039,ref045}. Descriptions of Gemma and Gemma 2, Mistral NeMo, and Granite 3.3 refer to their respective technical reports or official model resources \citep{ref013,ref014,ref021,ref032}. Information about the DeepSeek-V4 and Qwen-Max models and interfaces is drawn from a technical report and official documentation, respectively \citep{ref001,ref010}. These references describe the sources of the models; the specific versions evaluated are identified by their runtime model identifiers.

\textbf{Chunking-strategy ablation.} In addition to the four main benchmark configurations, a focused ablation compares RagFlow's paper chunking mode (RAP) with its general chunking mode (RAG) on a controlled subset of five documents per model and extraction task. The paired comparisons cover Qwen2.5-7B, DeepSeek-R1-14B, DeepSeek-R1-32B, and Dolphin3-8B, using the field-level precision, recall, and F1 definitions given above. Dolphin3-8B is included only in this auxiliary experiment, which is reported separately from the 15-model full-dataset comparison. Here, RAP and RAG identify chunking-mode variants and are distinct from the RAG-I and RAG-A knowledge-base configurations. For each metric $X\in\{P,R,F1\}$, the difference is $\Delta X=X_{\mathrm{RAP}}-X_{\mathrm{RAG}}$, expressed in percentage points. The results are reported in Table~\ref{tab:chunking-ablation}.

\begin{table}[tp]

\begingroup
\renewcommand{\baselinestretch}{1}\selectfont
\small

\hrule height 0.8pt
\vspace{3pt}

\noindent\textbf{Algorithm 1. RAGFlow-based Automated Evaluation Pipeline}

\vspace{3pt}
\hrule height 0.4pt
\vspace{3pt}

\noindent\textbf{Input:}
Paper collection $D=\{d_i\}_{i=1}^{N}$;
source-file sets $\{G_i\}_{i=1}^{N}$;
target-instance sets $\{M_i\}_{i=1}^{N}$;
gold standard $Y$;
parameter schema $S$;
candidate models $F$;
RAGFlow knowledge base $K$.

\vspace{1pt}

\noindent\textbf{Output:}
Document-level predictions $\hat{Y}$;
field-level evaluation records $E$;
aggregated model statistics $\mathcal{M}$.

\vspace{2pt}

\noindent
$\operatorname{ExtractAndAssemble}$ iterates over the targets in
$M_i$, using one CSV request per PSC target or three field-group
requests per nanozyme target (Fig.~\ref{fig:task-specific-prompts-psc});
$Q_i$ stores raw responses and $P_i$ stores assembled records.

\vspace{2pt}

\renewcommand{\arraystretch}{1}
\setlength{\tabcolsep}{0pt}

\begin{tabularx}{\linewidth}{
@{}r@{\hspace{1em}}
>{\raggedright\arraybackslash}X@{}
}

1 &
Initialize $\hat{Y}\leftarrow\varnothing$,
$E\leftarrow\varnothing$,
$\mathcal{M}\leftarrow\varnothing$. \\

2 &
\textbf{For each} model $f_{\theta}\in F$ \textbf{do} \\

3 &
\hspace*{1.2em}
$\hat{Y}_{\theta}\leftarrow\varnothing$,
$E_{\theta}\leftarrow\varnothing$;
$\operatorname{Clear}(K)$. \\

4 &
\hspace*{1.2em}
\textbf{For each} paper $d_i\in D$ \textbf{do} \\

5 &
\hspace*{2.4em}
$C_i\leftarrow
\operatorname{ParseAndChunk}
(\operatorname{Import}(G_i,K))$. \\

6 &
\hspace*{2.4em}
$R_i\leftarrow
\operatorname{Retrieve}(C_i,S,M_i)$. \\

7 &
\hspace*{2.4em}
$(Q_i,P_i)\leftarrow
\operatorname{ExtractAndAssemble}(f_{\theta},M_i,S,R_i)$;
$\operatorname{Save}(Q_i,P_i,d_i,M_i)$;
$\operatorname{Remove}(K,G_i)$. \\

8 &
\hspace*{2.4em}
$(A_i,U_i)\leftarrow
\operatorname{Match}(P_i,M_i)$. \\

9 &
\hspace*{2.4em}
\textbf{For each} target $m_{i,k}\in M_i$ \textbf{do} \\

10 &
\hspace*{3.6em}
$(p_{i,k},y'_{i,k})\leftarrow
\operatorname{Normalize}
(A_i[m_{i,k}],y_{i,k},S)$. \\

11 &
\hspace*{3.6em}
$e_{i,k}\leftarrow
\operatorname{Evaluate}(p_{i,k},y'_{i,k})$;
$\operatorname{Append}(\hat{Y}_{\theta},p_{i,k})$;
$\operatorname{Append}(E_{\theta},e_{i,k})$. \\

12 &
\hspace*{2.4em}
\textbf{End target loop}. \\

13 &
\hspace*{2.4em}
$\operatorname{AppendFP}(E_{\theta},U_i)$. \\

14 &
\hspace*{1.2em}
\textbf{End paper loop}. \\

15 &
\hspace*{1.2em}
$m_{\theta}\leftarrow
\operatorname{Aggregate}(E_{\theta})$. \\

16 &
\hspace*{1.2em}
$\operatorname{Append}(\hat{Y},\hat{Y}_{\theta})$;
$\operatorname{Append}(E,E_{\theta})$;
$\operatorname{Append}(\mathcal{M},m_{\theta})$. \\

17 &
\textbf{End model loop}. \\

18 &
\textbf{Return} $\hat{Y}$, $E$, $\mathcal{M}$. \\

\end{tabularx}

\vspace{3pt}
\hrule height 0.8pt

\endgroup

\end{table}
\section{Results and discussion}
\subsection{Overall benchmark performance}

Tables 3 and 4 report field-level precision, recall, and F1 for the nanozyme and PSC datasets, respectively, and Fig. 4 summarizes the distribution of model F1 scores across the four configurations. The following comparisons are based on the aggregated experimental results, with cross-model means computed using equal weights for the models.

In the RAG-I configuration for the nanozyme task, Qwen-Max and DeepSeek-V4 achieve F1 scores of 36.86\% and 35.78\%, respectively. Among locally deployed models, DeepSeek-R1-14B achieves the highest F1 (32.87\%), followed by Qwen2.5-14B (31.81\%) and Qwen2.5-7B (31.46\%). The performance differences between the main workflow and the control configurations vary across models. For Mistral-Nemo-12B, both RAG-I and RAG-A achieve a reported F1 score of 5.35\%. 
Thus, neither configuration demonstrates an F1 advantage for this model at the reported precision.

\begin{table}[tp]
\centering
\caption{Field-level evaluation results for the four extraction configurations on the nanozyme dataset (\%).}
\label{tab:3}

\begingroup
\scriptsize
\setlength{\tabcolsep}{0.8pt}
\renewcommand{\arraystretch}{1.12}

\begin{tabularx}{\linewidth}{@{}
l*{12}{>{\centering\arraybackslash}X}@{}}
\toprule
Model
& \multicolumn{3}{c}{RAG-I}
& \multicolumn{3}{c}{Direct-U}
& \multicolumn{3}{c}{RAG-A}
& \multicolumn{3}{c}{Direct-C}\\

\cmidrule(lr){2-4}
\cmidrule(lr){5-7}
\cmidrule(lr){8-10}
\cmidrule(lr){11-13}

& P & R & F1
& P & R & F1
& P & R & F1
& P & R & F1\\
\midrule

Qwen2.5-7B
& 58.23 & 21.55 & 31.46
& 0.22 & 4.46 & 0.41
& 27.60 & 14.42 & \textbf{18.94}
& 27.05 & 0.78 & 1.51\\

Llama3.1-8B
& 46.33 & 13.10 & 20.42
& 3.83 & 1.91 & 2.54
& 3.18 & 1.74 & 2.25
& 46.44 & 3.44 & 6.40\\

DeepSeek-R1-7B
& 47.21 & 16.30 & 24.24
& 1.90 & 0.27 & 0.48
& 3.70 & 2.68 & 3.10
& 24.81 & 0.16 & 0.31\\

DeepSeek-R1-14B
& 63.06 & 22.23 & 32.87
& 9.04 & 4.27 & 5.80
& 20.29 & 10.44 & 13.78
& 30.21 & 1.20 & 2.31\\

DeepSeek-R1-32B
& 65.15 & 17.96 & 28.16
& 9.46 & 9.78 & 9.62
& 23.65 & 7.04 & 10.85
& 12.24 & 0.37 & 0.72\\

Mistral-7B
& 45.47 & 12.98 & 20.20
& 4.13 & 1.36 & 2.04
& 5.51 & 2.90 & 3.80
& 24.90 & 2.77 & 4.99\\

Qwen2.5-14B
& 62.67 & 21.31 & 31.81
& 9.04 & 3.99 & 5.54
& 11.49 & 4.98 & 6.94
& 25.36 & 1.57 & 2.96\\

Llama3-8B
& 46.68 & 6.79 & 11.85
& 2.65 & 0.43 & 0.73
& 7.89 & 3.51 & 4.86
& 42.33 & 2.52 & 4.76\\

GLM4-9B
& 50.69 & 11.94 & 19.32
& 3.31 & 4.71 & 3.89
& 20.87 & 12.59 & 15.71
& 26.94 & 1.49 & 2.83\\

Granite3.3-8B
& 48.46 & 5.29 & 9.54
& 7.38 & 4.46 & 5.56
& 8.84 & 5.68 & 6.92
& 38.59 & 5.18 & 9.13\\

Gemma2-9B
& 42.88 & 4.57 & 8.25
& 9.71 & 0.37 & 0.72
& 41.66 & 0.13 & 0.26
& 34.20 & 2.65 & 4.91\\

Mistral-Nemo-12B
& 41.42 & 2.86 & 5.35
& 4.35 & 1.18 & 1.85
& 7.96 & 4.04 & 5.35
& 13.93 & 1.13 & 2.10\\

Gemma-7B
& 45.59 & 4.72 & 8.55
& 15.05 & 0.22 & 0.42
& 19.79 & 0.09 & 0.17
& 34.07 & 2.79 & 5.17\\

DeepSeek-V4
& 71.55 & 23.85 & 35.78
& 8.92 & 9.76 & 9.32
& 36.10 & 3.43 & 6.27
& 40.53 & 6.70 & \textbf{11.50}\\

Qwen-Max
& 74.81 & 24.46 & \textbf{36.86}
& 11.80 & 11.89 & \textbf{11.85}
& 43.20 & 7.09 & 12.17
& 35.71 & 3.82 & 6.90\\

\bottomrule
\end{tabularx}
\endgroup

\vspace{0.35em}

\begin{minipage}{\linewidth}
\footnotesize
\textit{Note:} Boldface indicates the highest F1 among individual models within each configuration. DeepSeek-V4 and Qwen-Max are accessed via APIs; the remaining 13 models are deployed locally through Ollama.
\end{minipage}
\end{table}

For the PSC task, RAG-I achieves a higher F1 than each respective control configuration for all 15 evaluated models. Qwen-Max and DeepSeek-V4 score 52.45\% and 49.56\%, respectively, while Qwen2.5-14B, DeepSeek-R1-32B, and Qwen2.5-7B achieve 42.22\%, 39.55\%, and 36.95\%. Together, the results across both domains indicate that RAG-I generally outperforms direct extraction and cumulative-knowledge-base configurations under the present evaluation conditions, although the magnitude of the advantage varies by model and materials system.

\begin{table}[tp]
\centering
\caption{Field-level evaluation results for the four extraction configurations on the PSC dataset (\%).}
\label{tab:4}

\begingroup
\scriptsize
\setlength{\tabcolsep}{0.8pt}
\renewcommand{\arraystretch}{1.12}

\begin{tabularx}{\linewidth}{@{}
l*{12}{>{\centering\arraybackslash}X}@{}}
\toprule
Model
& \multicolumn{3}{c}{RAG-I}
& \multicolumn{3}{c}{Direct-U}
& \multicolumn{3}{c}{RAG-A}
& \multicolumn{3}{c}{Direct-C}\\

\cmidrule(lr){2-4}
\cmidrule(lr){5-7}
\cmidrule(lr){8-10}
\cmidrule(lr){11-13}

& P & R & F1
& P & R & F1
& P & R & F1
& P & R & F1\\
\midrule

Qwen2.5-7B
& 42.26 & 32.82 & 36.95
& 0.29 & 0.45 & 0.35
& 1.89 & 0.90 & 1.22
& 9.10 & 0.54 & 1.02\\

Llama3.1-8B
& 27.11 & 26.66 & 26.89
& 0.07 & 6.79 & 0.15
& 0.21 & 0.32 & 0.25
& 20.39 & 4.77 & 7.73\\

DeepSeek-R1-7B
& 16.01 & 7.06 & 9.80
& 0.51 & 0.45 & 0.48
& 0.00 & 0.00 & 0.00
& 50.00 & 0.05 & 0.09\\

DeepSeek-R1-14B
& 30.37 & 16.46 & 21.35
& 3.08 & 4.86 & 3.77
& 2.80 & 2.43 & 2.60
& 16.90 & 3.82 & 6.23\\

DeepSeek-R1-32B
& 52.21 & 31.84 & 39.55
& 1.06 & 2.16 & 1.42
& 15.91 & 3.15 & \textbf{5.26}
& 28.45 & 2.97 & 5.38\\

Mistral-7B
& 34.94 & 30.35 & 32.48
& 0.37 & 0.76 & 0.50
& 1.79 & 1.80 & 1.79
& 27.13 & 4.59 & 7.85\\

Qwen2.5-14B
& 46.16 & 38.89 & 42.22
& 1.55 & 3.42 & 2.13
& 5.10 & 3.51 & 4.16
& 22.71 & 10.39 & 14.26\\

Llama3-8B
& 24.44 & 7.91 & 11.96
& 0.00 & 0.00 & 0.00
& 0.00 & 0.00 & 0.00
& 18.32 & 1.08 & 2.04\\

GLM4-9B
& 16.73 & 6.39 & 9.24
& 1.01 & 6.39 & 1.74
& 2.16 & 1.35 & 1.66
& 16.05 & 1.08 & 2.02\\

Granite3.3-8B
& 28.47 & 14.57 & 19.27
& 1.38 & 3.01 & 1.89
& 1.67 & 2.79 & 2.09
& 6.48 & 0.32 & 0.60\\

Gemma2-9B
& 25.85 & 10.57 & 15.00
& 1.07 & 0.58 & 0.76
& 0.20 & 0.54 & 0.29
& 14.19 & 1.84 & 3.26\\

Mistral-Nemo-12B
& 32.63 & 10.43 & 15.81
& 0.00 & 0.00 & 0.00
& 0.54 & 0.90 & 0.67
& 30.77 & 5.04 & 8.66\\

Gemma-7B
& 27.48 & 15.91 & 20.16
& 0.00 & 0.00 & 0.00
& 0.00 & 0.00 & 0.00
& 25.94 & 2.17 & 4.00\\

DeepSeek-V4
& 51.56 & 47.71 & 49.56
& 4.57 & 3.87 & 4.19
& 4.75 & 3.42 & 3.98
& 31.62 & 27.97 & \textbf{29.68}\\

Qwen-Max
& 59.14 & 47.12 & \textbf{52.45}
& 12.91 & 14.79 & \textbf{13.79}
& 4.21 & 1.71 & 2.43
& 55.54 & 15.56 & 24.31\\

\bottomrule
\end{tabularx}
\endgroup

\vspace{0.35em}

\begin{minipage}{\linewidth}
\footnotesize
\textit{Note:} P, R, and F1 denote precision, recall, and F1, respectively. RAG-I: paper-level isolated retrieval; Direct-U: direct extraction without predefined field names; RAG-A: cumulative-knowledge-base retrieval; Direct-C: direct extraction conditioned on paper and material identities. Model deployment modes are as in Table~\ref{tab:3}.
\end{minipage}
\end{table}

Table 5 reports cross-model mean metrics for each configuration. For the nanozyme task, the mean F1 scores for RAG-I, Direct-U, RAG-A, and Direct-C are 21.64\%, 4.05\%, 7.42\%, and 4.43\%, respectively; for the PSC task, the corresponding values are 26.85\%, 2.08\%, 1.76\%, and 7.81\%. These results characterize the overall performance of the evaluated model set. Mean F1 cannot be reconstructed from mean precision and mean recall.

\begin{table}[t]
\centering
\caption{Arithmetic means of metrics across the 15 evaluated models in the four configurations (\%).}
\label{tab:5}

\footnotesize
\renewcommand{\arraystretch}{1.12}

\setlength{\tabcolsep}{4pt}
\begin{tabularx}{\linewidth}{@{}
l>{\raggedright\arraybackslash}X
*{3}{>{\centering\arraybackslash}X}@{}}
\toprule
Dataset & Extraction configuration & Mean P & Mean R & Mean F1\\
\midrule
Nanozyme & RAG-I   & 54.01 & 13.99 & 21.64\\
          & Direct-U & 6.72 & 3.94 & 4.05\\
          & RAG-A    & 18.78 & 5.38 & 7.42\\
          & Direct-C & 30.49 & 2.44 & 4.43\\
PSC       & RAG-I    & 34.36 & 22.98 & 26.85\\
          & Direct-U & 1.86 & 3.17 & 2.08\\
          & RAG-A    & 2.75 & 1.52 & 1.76\\
          & Direct-C & 24.91 & 5.48 & 7.81\\
\bottomrule
\end{tabularx}
\end{table}

Model rankings change with extraction configuration. Qwen-Max achieves the highest F1 under both RAG-I and Direct-U in both domains. DeepSeek-V4 ranks first under Direct-C, with F1 scores of 11.50\% and 29.68\% in the nanozyme and PSC tasks, respectively. Under RAG-A, the highest F1 scores are obtained by Qwen2.5-7B for nanozymes (18.94\%) and DeepSeek-R1-32B for PSCs (5.26\%). Thus, model performance should be interpreted in the context of the actual evidence inputs and task constraints.

\begin{figure}[tp]
\centering
\includegraphics[
  width=\linewidth,
  height=.62\textheight,
  keepaspectratio
]{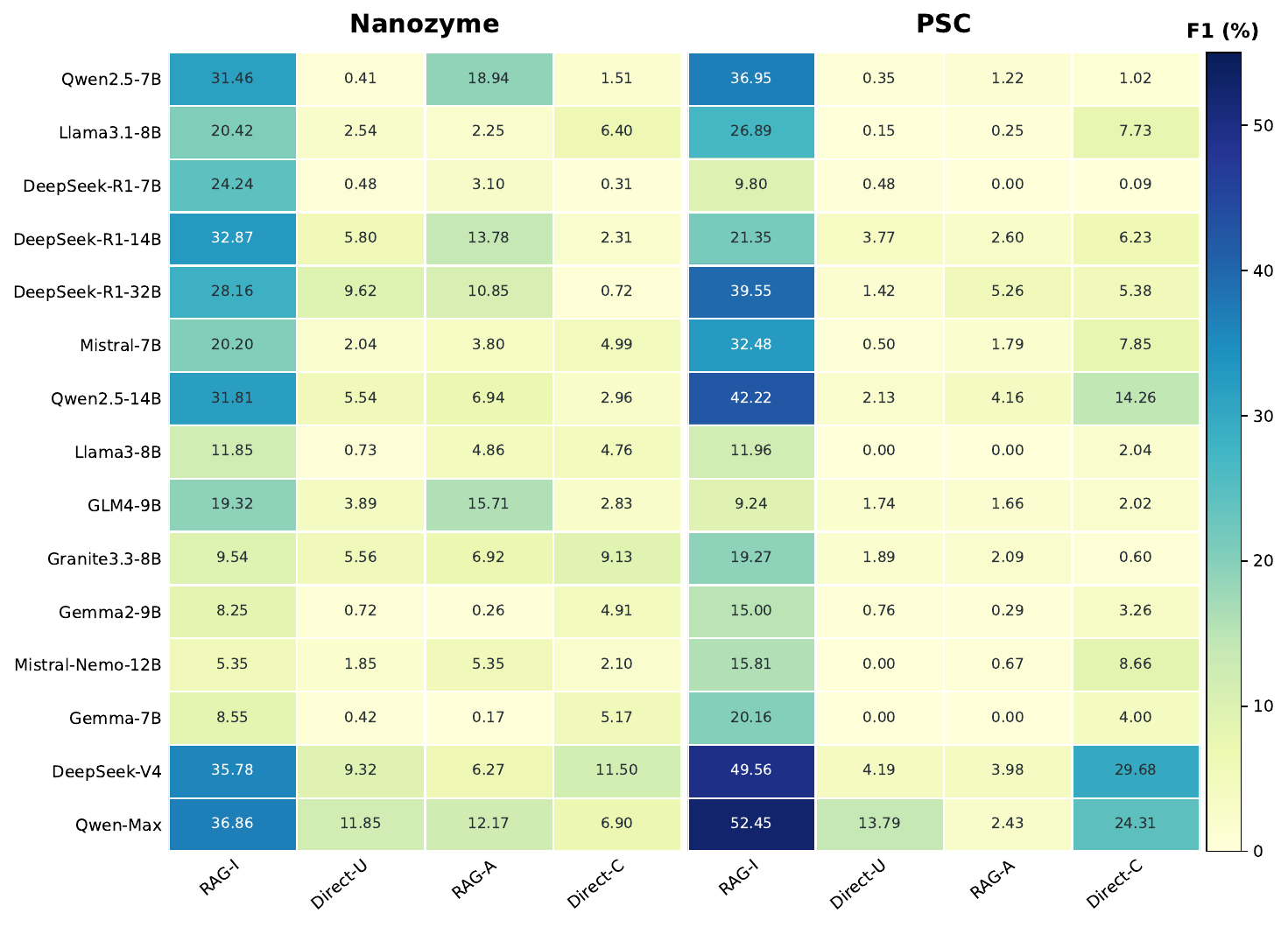}
\caption{F1 heatmap of models and extraction configurations on both datasets. Each cell shows the reported F1 (\%), with a shared color scale across the two domains.}
\label{fig:4}
\end{figure}
\subsection{Knowledge-base state and paper-level isolation}
The principal difference between RAG-I and RAG-A is whether previously processed papers are retained. RAG-I clears the knowledge base after extracting all targets in the current paper and saving the results, whereas RAG-A retains historical papers, so the knowledge base accumulates content during processing. In the PSC main workflow, the main article and its corresponding supplement are removed together to maintain the paper evidence group as the isolation boundary.

For the nanozyme task, RAG-I exceeds RAG-A in cross-model mean F1 by 14.22 percentage points. Qwen-Max improves from 12.17\% to 36.86\%, and DeepSeek-R1-14B from 13.78\% to 32.87\%; the two reported scores for Mistral-Nemo-12B are close. These results show that the magnitude of the response to the complete retrieval configuration varies by model.

For the PSC task, RAG-I exceeds RAG-A in mean F1 by 25.09 percentage points. Qwen-Max increases from 2.43\% to 52.45\%, DeepSeek-V4 from 3.98\% to 49.56\%, and Qwen2.5-14B from 4.16\% to 42.22\%. As shown in Fig. 5, the configuration-related differences are larger for most models in the PSC task.

These results indicate that knowledge-base retention should be explicitly reported as a condition of retrieval-augmented evaluation. Paper-level isolation aligns the scope of candidate evidence with the source of the records being scored. For papers with supplementary materials, this scope should encompass the entire evidence group. Results from the cumulative knowledge base must also be interpreted in light of paper-processing order and retrieval-source filtering conditions.

Aggregated field metrics are insufficient to identify specific error mechanisms. Retaining historical papers may expand the scope of candidate evidence. However, determining whether other papers were actually retrieved, whether material instances were confused, and how the model used the associated context requires retrieval provenance and field-level records. Therefore, the observed differences are not interpreted as direct proof of cross-document contamination.

\begin{figure}[tp]
\centering
\includegraphics[
  width=\linewidth,
  height=.62\textheight,
  keepaspectratio
]{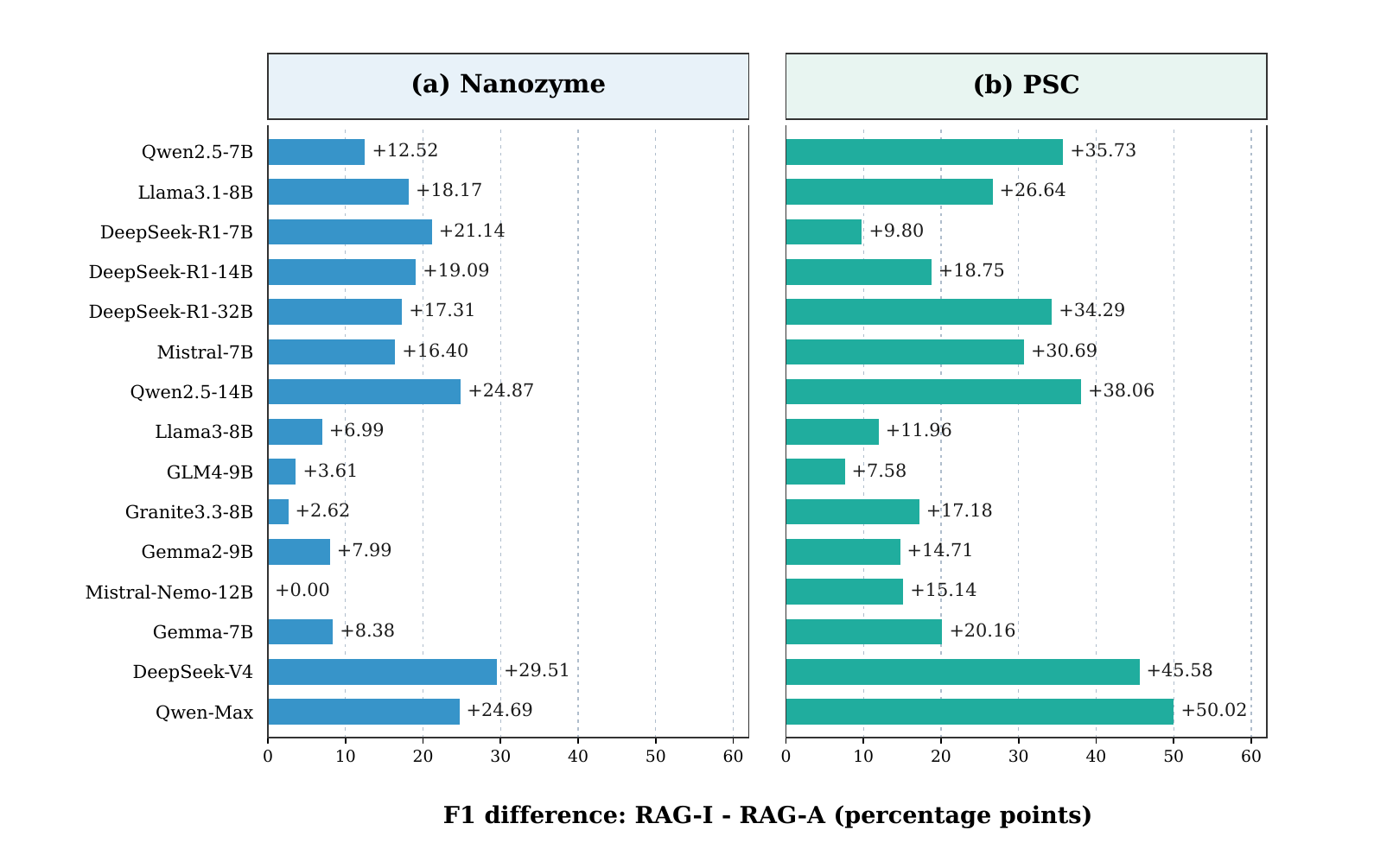}
\caption{Model-wise F1 differences between RAG-I and RAG-A on the (a) nanozyme and (b) PSC datasets. Positive values indicate higher F1 under RAG-I.}
\label{fig:5}
\end{figure}

\subsection{Input conditions for direct extraction}
Both Direct-U and Direct-C bypass RagFlow retrieval and directly invoke the corresponding local or API-based models on the paper input. Direct-U does not specify fixed parameter names, whereas Direct-C designates extraction targets using the paper ID and material name. The comparison concerns these two actual input configurations and does not assume that they differ in only one prompt element.

For the nanozyme task, relative to Direct-U, Direct-C increases mean precision from 6.72\% to 30.49\% while reducing mean recall from 3.94\% to 2.44\%; mean F1 increases only from 4.05\% to 4.43\%. Changes in F1 differ in direction across models: Llama3.1-8B improves from 2.54\% to 6.40\%, whereas DeepSeek-R1-32B decreases from 9.62\% to 0.72\% and Qwen-Max from 11.85\% to 6.90\%. Thus, providing target conditions in this configuration does not consistently improve field coverage.

For the PSC task, Direct-C yields a mean F1 of 7.81\%, exceeding Direct-U at 2.08\%, and 13 of the 15 models obtain higher F1 scores. DeepSeek-V4 increases from 4.19\% to 29.68\%, Qwen2.5-14B from 2.13\% to 14.26\%, and Qwen-Max from 13.79\% to 24.31\%. The different patterns across the two tasks suggest that the effects of input constraints must be interpreted in the context of the parameter schema and evaluated model.

Compared with Direct-U, RAG-I improves mean F1 by 17.59 and 24.77 percentage points on nanozymes and PSCs, respectively; compared with Direct-C, the improvements are 17.21 and 19.04 percentage points. These differences correspond to the complete workflow combining paper-level evidence isolation, construction of retrieval context, field-schema constraints, and the task-specific query and output-assembly procedures.

The four configurations do not form a fully orthogonal design of retrieval, target conditions, and field schemas. Accordingly, the observed differences cannot estimate the independent contribution of any single component. Further mechanistic analysis would require holding other conditions fixed while separately controlling evidence retrieval and prompt constraints.

\subsection{Ablation study of chunking strategies}
\label{sec:chunking-ablation}
Table~\ref{tab:chunking-ablation} reports the paired results on the controlled five-document subsets. Paper chunking achieves higher F1 than general chunking for all four models in both extraction tasks. For nanozymes, F1 gains range from 8.06 to 28.04 percentage points. DeepSeek-R1-32B shows the largest gain, increasing from 1.59\% to 29.63\%, while Qwen2.5-7B increases from 16.00\% to 35.23\%. For PSCs, F1 gains range from 2.50 to 17.58 percentage points. The general mode yields F1 scores of 0.00\% for both DeepSeek-R1 models on these subsets, whereas the paper mode yields 9.41\% for DeepSeek-R1-14B and 17.58\% for DeepSeek-R1-32B. These comparisons show that chunking-mode selection can materially affect extraction performance within the evaluated settings.

The gains are not uniform across all metrics. For Dolphin3-8B on nanozymes, precision increases by 18.82 percentage points, but recall decreases by 6.25 percentage points; F1 nevertheless improves from 9.77\% to 25.22\%. Thus, higher F1 does not imply simultaneous gains in precision and recall. Overall, the subset results favor paper chunking, but the five-document evaluation does not establish full-corpus generalization or statistical significance. This ablation addresses chunking mode separately from the main comparisons of knowledge-base state and direct-extraction inputs.

\begin{table}[t]
\centering
\caption{Chunking-strategy ablation on a controlled five-document subset per model and task.}
\label{tab:chunking-ablation}
\begingroup
\scriptsize
\renewcommand{\arraystretch}{1.12}
\setlength{\tabcolsep}{1pt}
\begin{tabularx}{\linewidth}{@{}lc*{12}{>{\centering\arraybackslash}X}@{}}
\toprule
Model & Mode
& \multicolumn{6}{c}{Nanozyme}
& \multicolumn{6}{c}{PSC}\\
\cmidrule(lr){3-8}\cmidrule(lr){9-14}
& & P & R & F1 & $\Delta P$ & $\Delta R$ & $\Delta F1$
  & P & R & F1 & $\Delta P$ & $\Delta R$ & $\Delta F1$\\
\midrule
Qwen2.5-7B & RAP
& 40.26 & 31.31 & \textbf{35.23} & 8.68 & 20.60 & 19.23
& 11.11 & 3.17 & \textbf{4.94} & 6.35 & 1.53 & 2.50\\
& RAG
& 31.58 & 10.71 & 16.00 & -- & -- & --
& 4.76 & 1.64 & 2.44 & -- & -- & --\\
\addlinespace[3pt]
DeepSeek-R1-14B & RAP
& 60.66 & 31.62 & \textbf{41.57} & 18.55 & 3.79 & 8.06
& 9.30 & 9.52 & \textbf{9.41} & 9.30 & 9.52 & 9.41\\
& RAG
& 42.11 & 27.83 & 33.51 & -- & -- & --
& 0.00 & 0.00 & 0.00 & -- & -- & --\\
\addlinespace[3pt]
DeepSeek-R1-32B & RAP
& 57.14 & 20.00 & \textbf{29.63} & 7.14 & 19.19 & 28.04
& 21.62 & 14.81 & \textbf{17.58} & 21.62 & 14.81 & 17.58\\
& RAG
& 50.00 & 0.81 & 1.59 & -- & -- & --
& 0.00 & 0.00 & 0.00 & -- & -- & --\\
\addlinespace[3pt]
Dolphin3-8B & RAP
& 24.58 & 25.89 & \textbf{25.22} & 18.82 & $-6.25$ & 15.45
& 7.69 & 5.45 & \textbf{6.38} & 4.99 & 3.56 & 4.16\\
& RAG
& 5.76 & 32.14 & 9.77 & -- & -- & --
& 2.70 & 1.89 & 2.22 & -- & -- & --\\
\bottomrule
\end{tabularx}
\endgroup

\vspace{0.35em}
\begin{minipage}{\linewidth}
\footnotesize
\textit{Note:} P, R, and F1 are percentages. RAP and RAG denote RagFlow's paper and general chunking modes, respectively. Differences are calculated as RAP minus RAG and reported in percentage points; negative values indicate a decrease. Dashes mark the baseline rows, for which differences are not applicable. Boldface denotes the higher F1 within each model--task pair. These subset results are separate from the full-dataset results in Tables~\ref{tab:3} and~\ref{tab:4}.
\end{minipage}
\end{table}

\subsection{Cross-domain comparison of materials systems}
The mean F1 of RAG-I is 26.85\% on PSCs, higher than the 21.64\% on nanozymes; the highest individual-model F1 scores are 52.45\% and 36.86\%, respectively. Twelve of the 15 models attain a higher F1 on PSCs. Figure 6 presents the highest individual-model scores for all four configurations, showing that the relationship between domains depends on the configuration.

The tasks also exhibit different precision--recall profiles. In the nanozyme main workflow, mean precision and recall are 54.01\% and 13.99\%, respectively, compared with 34.36\% and 22.98\% for PSCs. All 15 models have lower precision on PSCs, but higher recall yields higher F1 scores for most models. Cross-domain comparisons therefore need to consider both the correctness of predictions and coverage of reference fields.

These differences should be interpreted in relation to the field sets and evidence organization of the tasks. Nanozyme records contain kinetic parameters dependent on substrate and reaction conditions, whereas PSC records emphasize process variables, interface locations, and photovoltaic performance, with evidence jointly obtained from the main article and its supplementary material. Both tasks specify one target per request through \texttt{\{name\}}, but their query procedures differ: PSCs use one CSV request per target, while nanozymes use three sequential field-group queries per target. Because the field types, reporting densities, evidence structures, and query procedures differ, aggregate scores alone cannot establish that one materials domain is universally easier or more difficult.

Within-model-family scaling effects are likewise task dependent. In the nanozyme main workflow, DeepSeek-R1 models with 7B, 14B, and 32B parameters attain F1 scores of 24.24\%, 32.87\%, and 28.16\%, respectively; on PSCs, their scores are 9.80\%, 21.35\%, and 39.55\%. Increasing Qwen2.5 from 7B to 14B improves F1 by 0.35 percentage points on nanozymes and 5.27 percentage points on PSCs. These results indicate that the relationship between parameter scale and performance should be examined for specific model versions and materials tasks.

The experiments in the two domains constitute parallel evaluations rather than cross-domain transfer-learning experiments. Further explanation of changes in model ranking requires stratified analyses by field category, evidence location, and conditional dependence.

\begin{figure}[tp]
\centering
\includegraphics[width=\linewidth,height=.62\textheight,keepaspectratio]{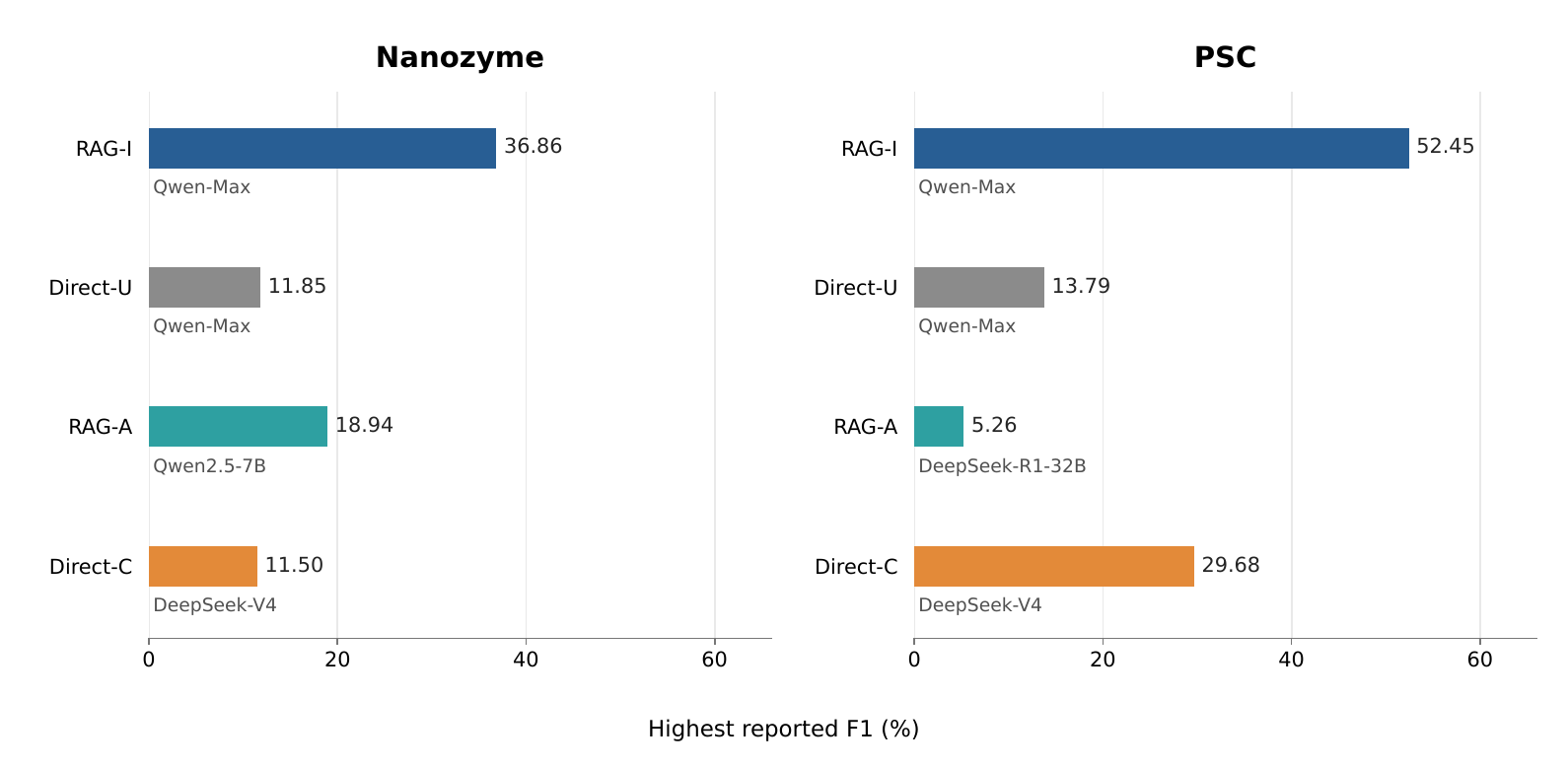}
\caption{Highest individual-model F1 for the four extraction configurations on both datasets. For each configuration, the highest reported score is selected from the same set of 15 candidate models.}
\label{fig:6}
\end{figure}

\subsection{Precision and recall analysis}
Examining precision and recall jointly distinguishes predictive characteristics that may be obscured by similar F1 scores. In the nanozyme main workflow, Qwen-Max attains precision and recall of 74.81\% and 24.46\%, respectively; the corresponding values for DeepSeek-V4 are 71.55\% and 23.85\%. Although both models rank highly in F1, many nonempty reference fields remain unrecovered, indicating that field coverage is a major limitation under the current setting.

High precision may also coexist with extremely low recall. For example, in nanozyme RAG-A, Gemma2-9B achieves precision of 41.66\% but recall of only 0.13\%, yielding F1 of 0.26\%. In PSC Direct-C, DeepSeek-R1-7B has precision of 50.00\% but recall of only 0.05\%. These results indicate limited coverage of reference fields by correct predictions; they do not establish that the models employ stable confidence-based selection strategies.

The leading models in the PSC main workflow show more balanced precision and recall. Qwen-Max records 59.14\% and 47.12\%, respectively, compared with 51.56\% and 47.71\% for DeepSeek-V4. Although Qwen-Max has slightly lower recall, its higher precision leads to a higher F1, indicating that the correctness of nonempty outputs and field coverage jointly determine the final score.

Conversely, low precision limits overall direct-extraction performance. In PSC Direct-U, Llama3.1-8B achieves recall of 6.79\% but precision of only 0.07\%. Qwen2.5-7B records 0.45\% recall and 0.29\% precision. Generating parameter-related content is not sufficient to produce valid predictions consistent with the target instance, field schema, and gold standard. Specific errors require inspection of output parsing, instance matching, and field-comparison records.

Figure 7 summarizes the precision--recall distributions for all four configurations. Changes in a model's position across configurations indicate that F1 improvements may result from greater prediction correctness, improved field coverage, or both; no single metric sufficiently describes parameter recovery quality.

Accordingly, precision, recall, and F1 are all reported. Retrieval omissions, instance mismatches, unit inconsistencies, and output-parsing failures are identified as potential error sources requiring further record-level analysis.

\begin{figure}[tp]
\centering
\includegraphics[width=\linewidth,height=.62\textheight,keepaspectratio]{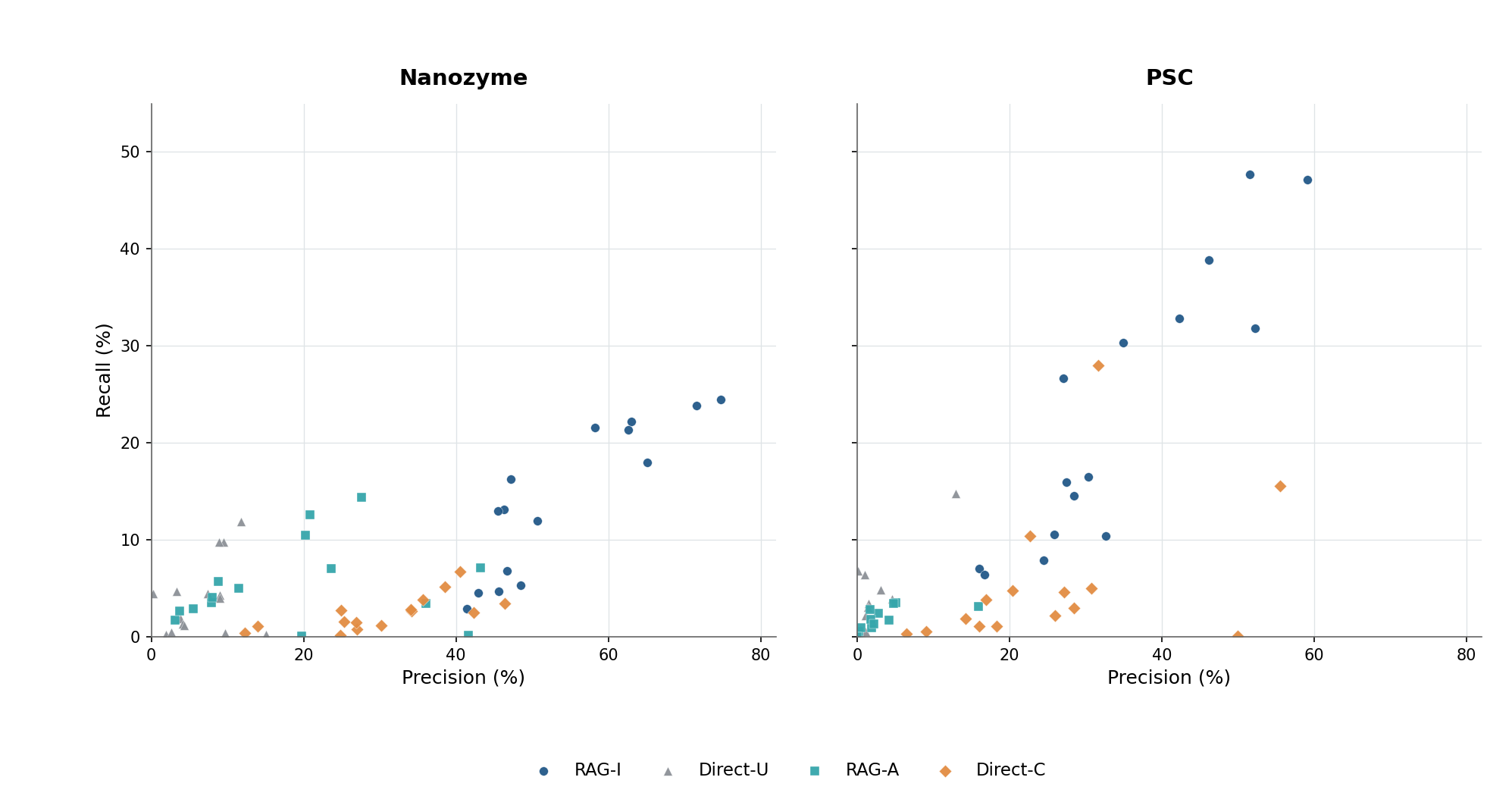}
\caption{Precision--recall distributions under the four configurations. The left panel shows nanozymes and the right panel PSCs; colors and markers distinguish the extraction configurations.}
\label{fig:7}
\end{figure}
\subsection{Discussion and limitations}
The four experimental configurations indicate that parameter recovery from materials literature should be evaluated in terms of both the model and the extraction workflow. The paper-level isolated main workflow generally achieves higher F1, while changes in knowledge-base retention and input conditions for direct extraction are associated with performance differences of varying magnitude. Model results should therefore be reported with explicit descriptions of evidence sources, target conditions, field constraints, and the knowledge-base lifecycle so that each score corresponds to an interpretable experimental setting.

For studies accompanied by supplementary materials, paper-level isolation must also account for evidence completeness. The PSC main text and supplementary file constitute evidence from the same study. Joint retrieval preserves within-paper cross-file associations, while clearing the entire group constrains the scope of candidate evidence for subsequent papers. This process provides an explicit operational boundary for organizing and evaluating multi-file scientific literature.

This study has several limitations. First, the four main benchmark configurations are not fully orthogonal, so the independent effects of retrieval, prompt constraints, and knowledge-base state cannot be separated. The three-query nanozyme strategy also lacks a matched single-query control, and its independent contribution is not quantified. Second, aggregate metrics do not identify the individual contributions of error types arising during retrieval and generation, and the current results provide no uncertainty estimates based on repeated runs. Third, the two datasets differ in field and evidence structures, so cross-domain scores cannot replace task-difficulty analysis. The PSC experiments also lack a main-text-only control and thus cannot quantify the independent contribution of supplementary materials. The nanozyme prompt requires binary elemental outputs, while the evidence-based gold standard permits indeterminate elemental labels; this constraint can produce nonempty predictions for null reference fields and should be considered when interpreting elemental-field errors. Moreover, although expert verification helps correct candidate labels, it cannot, by itself, eliminate potential biases introduced by the model used for assisted annotation.

Future research could extend the chunking ablation to larger document subsets and compare source-constrained retrieval and main-text--supplementary-evidence integration while holding target sets, field schemas, and scoring rules fixed. Independent annotation review, field-level error analysis, and repeated-run estimates could further assess benchmark reliability. For nanozymes, attention should focus on evidence coverage for condition-dependent parameters; for PSCs, on correspondences among chemical components, application locations, and device performance.

\section{Conclusions}
This study introduces MatExtractBench, an entity-conditioned document-level parameter extraction benchmark covering nanozymes and PSCs. The RagFlow-based main workflow uses a single PDF for each nanozyme paper and a main-text--supplementary evidence group for each PSC paper. After extracting all targets in the current paper and saving the results, the associated knowledge-base content is cleared. Evaluation aligns predictions by predefined instances and measures structured parameter recovery using consistent field normalization and value-level counting rules.

The comparison of 13 locally deployed models and two API models under four configurations shows that the paper-level isolated workflow generally achieves higher F1. Qwen-Max obtains F1 scores of 36.86\% and 52.45\% on nanozymes and PSCs, respectively; compared with the cumulative-knowledge-base configuration, the main workflow improves mean F1 by 14.22 and 25.09 percentage points. The auxiliary five-document ablation also yields higher F1 with paper chunking for all four evaluated models in both tasks. The results also show that field coverage and instance attribution across different materials parameter structures require further improvement. MatExtractBench provides a unified data and protocol foundation for further research on dispersed-evidence integration, condition-dependent parameter extraction, and field-level evaluation in scientific literature.

\section*{CRediT authorship contribution statement}

\textbf{Jian He}: Conceptualization, Methodology, Software, Formal analysis, Investigation, Visualization, Writing -- original draft. \textbf{Yao Lu}: Methodology, Validation, Writing -- review \& editing. \textbf{Xin Cheng}: Software, Validation, Data curation. \textbf{Chenxiao Liu}: Investigation, Data curation, Validation. \textbf{Tianyi Jiang}: Investigation, Data curation, Validation. \textbf{Feng Xu}: Resources, Validation, Writing -- review \& editing. \textbf{Ying Li}: Resources, Validation, Writing -- review \& editing. \textbf{Qi Xin}: Investigation, Resources, Validation. \textbf{Yi Dong}: Resources, Validation, Writing -- review \& editing. \textbf{Peng Liu}: Resources, Validation, Writing -- review \& editing. \textbf{Shanqing Yu}: Conceptualization, Methodology, Supervision, Writing -- review \& editing. \textbf{Qi Xuan}: Supervision, Project administration, Writing -- review \& editing.

\section*{Declaration of competing interests}
The authors declare that they have no known competing 
financial interests or personal relationships that could 
have appeared to influence the work reported in this paper.

\section*{Acknowledgements}
This research was partially supported by the National Key Research and Development Program of China (2025YFA1510900), the Joint Fund of the Natural Science Foundation of Zhejiang Province and Baima Lake Laboratory (LBMHZ25F020002), and the Research Program of the State Key Laboratory of Green Chemical Synthesis and Conversion (SKLGSC-ZZ-202604). The authors thank the State Key Laboratory of Green Chemical Synthesis and Conversion and the Zhejiang Provincial Key Laboratory of Catalyst Surface and Interface Science and Engineering at Zhejiang University of Technology for their support.
\section*{Data and software availability statement}
The code and datasets used in this study are available at: \url{https://github.com/205jindadiao-wq/MatExtractBench}.

\section*{Declaration of generative AI and AI-assisted 
technologies in the manuscript preparation process}

During the preparation of this work, the authors used 
ChatGPT (OpenAI) to assist with translation, language 
editing, and improvement of readability. The authors 
reviewed and edited the output as needed and take full 
responsibility for the content of the published article.

\end{document}